\documentclass{elsart}
\usepackage{natbib,psfig}
\begin{document}
\runauthor{Carilli}
\begin{frontmatter}
\title{Xray observations of high redshift radio galaxies}
\author[chrisc]{C.L. Carilli}

\address[chrisc]{NRAO, Socorro, NM, USA}

\begin{abstract}

I summarize Xray properties of high redshift radio galaxies, beginning
with a brief review of what has been learned from Xray observations of
low redshift powerful radio galaxies (in particular, Cygnus A), and
then turning to Chandra observations of four high redshift radio
galaxies. Hot Xray emitting atmospheres of the type seen in low
redshift clusters are not detected in the high redshift sources,
suggesting that these systems are not yet virialized massive clusters,
but will likely evolve into such.  Xray emission from highly obscured
AGN is detected in all cases.  Extended Xray emission is also seen,
and the extended emission is clearly aligned with the radio
source, and on a similar spatial scale. Multiple mechanisms are
proposed for this radio-Xray alignment, including inverse Compton
scattering of photons from the AGN (the 'Brunetti mechanism'), and
thermal emission from ambient gas that is shocked heated by the
expanding radio source. The pressure in the high filling factor
shocked gas is adequate to confine the radio source and the low
filling factor optical line emitting clouds.

\end{abstract}

\end{frontmatter}

\section{Introduction}

There are a number of possible Xray emission mechanisms for high
redshift radio galaxies.  Thermal emission from hot gas probes the
large scale gas distribution and the hydrodynamic interaction between
the expanding radio source and this gas. Thermal and/or non-thermal
emission from the AGN, and soft Xray absorption toward the AGN, can be
used to study physical processes in the AGN and conditions in the
local AGN environment.  Non-thermal emission from radio jets
(synchrotron and/or inverse Compton (IC)) can be used to constrain the
relativistic particle acceleration and loss mechanisms, and the source
magnetic fields. Most recently, detection of Xray emission from
(likely) cluster members in one high z radio galaxy has revealed the
(obscured) AGN population in the associated proto-cluster. 

This paper summarizes what has been learned from Xray observations of
lower redshift radio galaxies, and then reviews observations of higher
redshift sources.  Particular attention is paid to the source
PKS1138-262 at z=2.2, which shows evidence for most of the phenomena
discussed above, and to the radio-Xray 'alignment' seen in this, and
other, high redshift radio galaxies. I use H$_o = 65$ km s$^{-1}$
Mpc$^{-1}$ and a flat cosmology with $\rm \Omega_m = 0.3$ throughout.


\section{Xray observations of low redshift radio galaxies}

The high redshift radio galaxies discussed below are all
very high luminosity sources, with P$_{\rm 178 MHz} > 5\times10^{28}$
W Hz$^{-1}$. The only radio source at $z < 0.4$ with comparable
luminosity is Cygnus A at $z = 0.057$ (Carilli \& Barthel 1996). 
Cygnus A has been extensively studied in the Xray, and
displays most of the thermal and nonthermal emission
phenomena discussed above (Harris et al. 1994; Carilli et al. 1994).
In this section I will use the recent high resolution Chandra images
of Cygnus A (Wilson et al. 2000; Smith et al. 2002; Young et al.
2002) to illustrate these various phenomena. 

\subsection{Thermal emission and hydrodynamics}

The Cygnus A radio source is situated at the center of a dense cluster
atmosphere that extends to a radius of at least 0.5 Mpc (Smith et
al. 2002). The total Xray luminosity of the cluster is L$_{\rm 2 - 10
keV} = 1.4\times10^{45}$ erg s$^{-1}$, at an average temperature of 8
keV, and the electron density at the radio hot spot radius (70 kpc) =
0.006 cm$^{-3}$ (Ueno et al. 1994; Reynolds \& Fabian 1996). The total
gas mass is $2\times10^{13}$ M$_\odot$ and the total gravitational
mass is $2\times10^{14}$ M$_\odot$ (Reynolds \& Fabian 1996).

The hydrodynamic evolution of a light, hypersonic jet propagating
into a hot intercluster medium (ICM) has been considered in detail by
many authors (see the contribution by Bicknell in these proceedings). 
The standard model involves a double shock structure, 
with a terminal jet shock and a stand-off, or bow, shock
propagating into the external medium (see Figure 1A). The two shocked
fluids   (jet and external) meet in pressure equilibrium along a
contact discontinuity.  The radio emission is thought to be
from shocked jet material contained within the contact discontinuity.

\begin{figure}
\psfig{figure=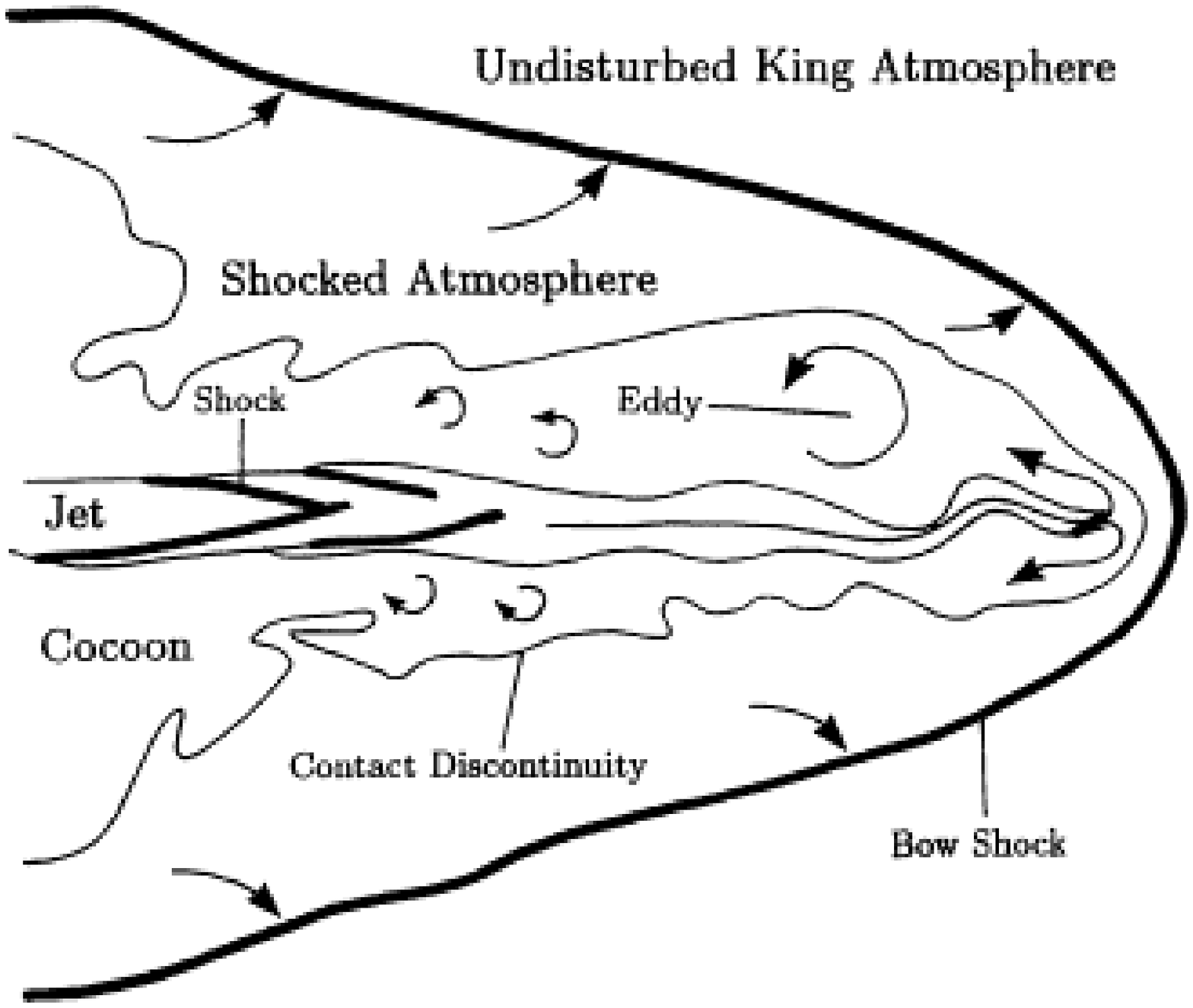,width=2.5in}
\vskip -2.5in
\hskip 3in
\psfig{figure=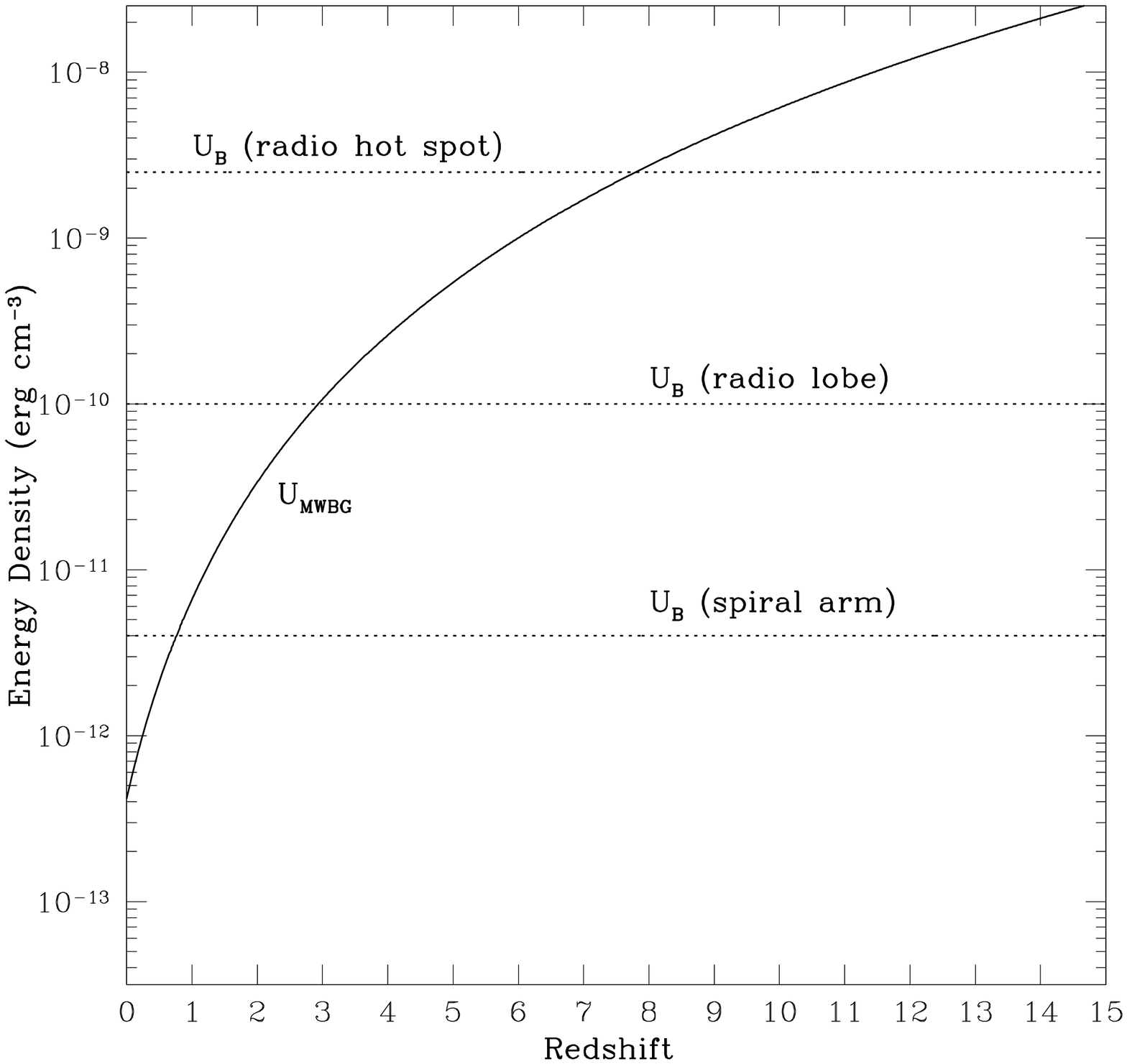,width=2.5in}
\caption{A. The figure on the left shows a heuristic model for
the hydrodynamic evolution of a light, hypersonic jet (Clarke
et al. 1997). B. The figure on the right show the
evolution with redshift of the
energy densities in the CMB compared to the magnetic energy
densities in different cosmic structures.  }
\end{figure}

\begin{figure}
\psfig{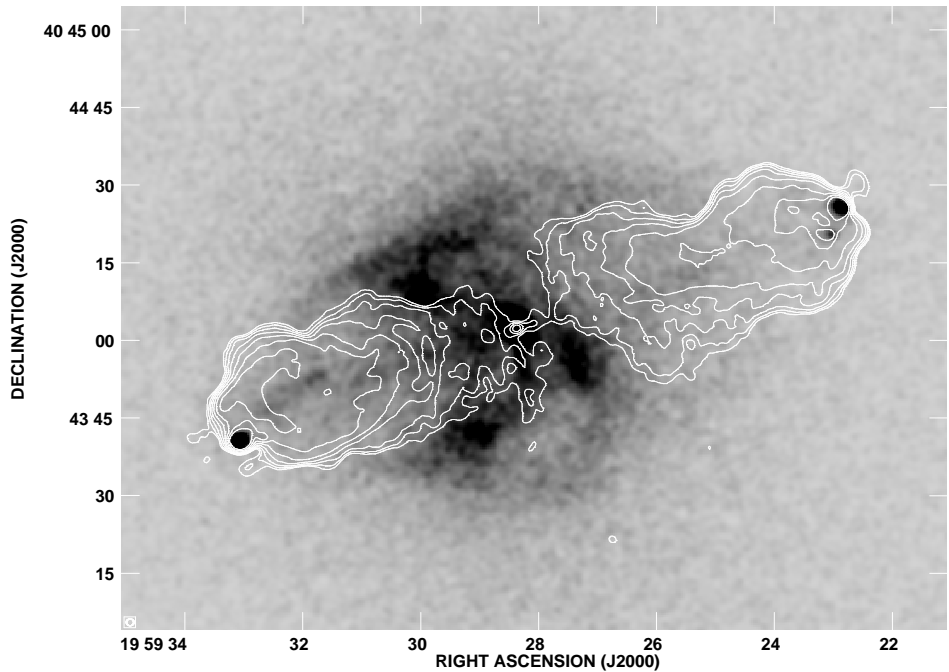}
\caption{The grey scale shows total Xray emission from the inner 200 kpc
of the Cygnus A cluster as seen by Chandra (Smith et al. 2002). The
contours show the VLA image of the radio source at 1.4 GHz.}
\end{figure}

A number of signatures in Xray images are expected from such a system,
including emission from the unperturbed atmosphere (see above), plus
excess emission from the shocked ICM surrounding the radio lobes, and
a deficit of emission from the evacuated radio lobes themselves. A
critical aspect of this process is the temperature dependence of Xray
emission, ie. the detected Xray emission will depend strongly on the
observed energy band, the temperature of the external medium, and the
strength and evolution of the bow shock (Clarke et al. 1997).

Figure 2 shows the radio image of Cygnus A plus the Chandra image of
the inner 200 kpc of the Cygnus A cluster (Smith et al. 2002). This
image shows clearly the interaction of the radio source and the
cluster gas, including deficits of Xray emission at the positions of
the radio lobes, and excesses of emission directly surrounding the
radio lobes.

As a probe of thermal processes, Xray observations provide direct
constraints on physical conditions in the emitting regions.  In
contrast, observations of nonthermal radio synchrotron radiation
provide only indirect physical probes requiring many assumptions
(eg. minimum energy). Smith et al. (2002) have considered the pressure
in the shocked thermal material enveloping the radio lobes in Cygnus
A, for which they derive a pressure of $\sim 1\times10^{10}$ dyne
cm$^{-2}$. For comparison, the minimum pressure in the radio lobes
derived from the synchrotron surface brightness is $2\times10^{11}$
dyne cm$^{-2}$. Given that the shocked fluids must be in pressure
equilibrium along the contact discontinuity, this difference suggests
a departure from minimum energy conditions in the radio lobes by a
factor five.

\subsection{Nonthermal emission}

Figure 2 also reveals Xray emission coincident with the radio hot
spots in Cygnus A. The spectral analysis of Wilson et al (2000) shows
that this Xray emission is consistent with nonthermal IC
emission from the same population of relativistic electrons emitting
radio synchrotron radiation.  In this case the dominant radiation
field is the radio synchrotron photons themselves, such that the
emission has been called synchrotron self Compton (SSC) emission
(Harris et al. 1994).  

Given a photon field, the IC Xray emissivity constrains the number
density of relativistic electrons, while the radio emissivity is a
function of the relativistic electron density and the magnetic field
strength. A comparison of the two then provides an estimate of the
magnetic field strength.  Wilson et al. (2002) derive fields of 150
$\mu$G for the radio hot spots based on the SSC and synchrotron
emissivities.  The minimum energy fields in the hot spots are about
250 $\mu$G, suggesting that the physical conditions in the hot spots
are not far from minimum energy.

An important point to keep in mind in this regard is the increase in
the energy density of the microwave background with redshift ($\rm
U_{CMB} = 4.2\times10^{-13}(1+z)^4$ erg cm$^{-3}$). Figure 1B shows
the evolution of $\rm U_{CMB}$ with redshift versus the typical
magnetic energy densities in galaxies and radio galaxies. The
implication is that in the disks of normal galaxies IC losses off the
CMB will dominate over synchrotron losses for galaxies beyond $z \sim
1$. The cross over point for radio lobes is $z \sim 3$, while that for
radio hot spots is $z \sim 7$.

The increasing significance of IC losses with redshift has led some to
hypothesize the possibility of IC-loud, but radio quite, jets at high
redshift (Schwartz et al. 2002). The important point is that the
spectral peak of the CMB behaves as $1.6\times10^{11}$(1+z) Hz, such
that observations at 1 keV are sensitive to electrons with $\gamma_e
\sim 1000$, independent of redshift. The radiative lifetime of such
electrons at $z \sim 6$ is about $1\times10^7$ years. For comparison,
observations of a $z=6$ jet at 1.4 GHz probe electrons with $\gamma_e
\sim 7000$, corresponding to radiative lifetimes of $2.5\times10^6$
years.

\subsection{Xrays from the obscured AGN}

Cygnus A shows a hard, highly obscured nuclear Xray source with N(HI)
= $2\times10^{23}$ cm$^{-2}$, $\Gamma = 1.5$, and $\rm L_{2 - 10 keV}
= 1.2\times10^{45}$ erg s$^{-1}$ (Young et al. 2002). Young et
al. (2002) also find evidence for a scattered Xray component with a
luminosity of about 1$\%$ of the nuclear Xray source. The scatted
component is co-spatial with the high ionization optical line emitting
regions, supporting the idea that these regions are photo-ionized by
radiation from the AGN.

As a matter of completeness, there have been Chandra observations
of a number of other luminous radio galaxies at $z \sim 0.5$
(Harris et al. 2002; Hardcastle et al. 2002). These observations
reveal similar phenomena as those seen in Cygnus A, although
not in such exquisite detail.

\section{Xray observations of high redshift radio galaxies}

\subsection{Cluster atmospheres and AGN emission}

Four high redshift powerful radio galaxies have been observed with
Chandra to date: 1138--262 at z=2.2, 0236--254 at z = 2.0, 3C 294 at z
= 1.8, and 0902+343 at z=3.4. All were detected, and all but 0902+343
show spatially extended Xray emission (Figure 3; for 3C 294 see
Crawford this volume; for 0902+343 see Fabian et al. 2002). The
emission mechanisms appear to span the full range of thermal and
non-thermal processes seen at low redshift, and I will consider each
in turn.

One of the primary drivers for observing high redshift radio galaxies
in the Xray was to search for large scale structure (ie. hot cluster
atmospheres) at high redshift, with the potential to place stringent
constraints on $\rm \Omega_M$ (Fabian et al. 2001).  The sources listed
above were all chosen because they showed clear evidence for being
situated in dense cosmic environments (large Ly $\alpha$ halos and
galaxy overdensities, extreme rotation measures, highly disturbed
radio morphologies, etc...).  

While three of the sources show
spatially extended Xray emission, in no case do we find evidence for a
massive cluster atmosphere. Typical limits to the luminosity of
cluster atmospheres associated with these sources are $\rm L_{2 -
10keV} < 2\times10^{44}$ erg s$^{-1}$.  

Considering clustering, in one source (1138--262), Pentericci et
al. (2002) have found an excess of Xray loud AGN in vicinity of the
radio source, by a factor two relative to the field. Six of these
sources are spectroscopically or photometrically confirmed to be at
the redshift of the radio galaxy.

Xray emission is detected from the nuclei of all the sources observed
thus far.  The data are consistent with highly obscured powerlaw
spectra with: $\rm L_{2 - 10 keV} \sim 2\times 10^{45}$ erg s$^{-1}$,
$\Gamma \sim -1.5$, and N(HI) $\sim 1\times10^{-23}$. The implied
visual extinction is about $\rm A_V \sim 60$ assuming a Galactic
dust-to-gas ratio. The ratio of the core radio to Xray luminosities
for these sources are within the range defined by steep spectrum radio
loud quasars. Using the relationship between core Xray luminosity and
bolometric QSO luminosity, the implied bolometric luminosities for
these AGN are between 10$^{46}$ to 10$^{47}$ erg s$^{-1}$ (Brinkmann
et al. 1997).

\subsection{Radio-Xray alignment effect}

The most surprising aspect of the Xray observations of high redshift
radio galaxies is that the extended Xray emission is closely aligned
with the radio axis, and on a similar scale (Figure 3; 
see also Crawford this volume). Like the
radio-optical alignment effect for high z radio galaxies, it appears
that multiple mechanisms may be responsible for the radio-Xray
alignment effect.

One possible mechanism is the West/Barthel-Arnaud effect (West 2000;
Barthel \& Arnaud 2000). The basic idea is that galaxies form along
filamentary structures, and that radio jets propagating along the axis
of highest density have higher conversion efficiency of jet kinetic
energy into radio luminosity.  The original thermal interpretation of
the Xray emission in 3C 294 by Fabian et al. (2001) would be
consistent with this model, although more recent data has called this
original interpretation into question (see Crawford this volume).

\begin{figure}
\psfig{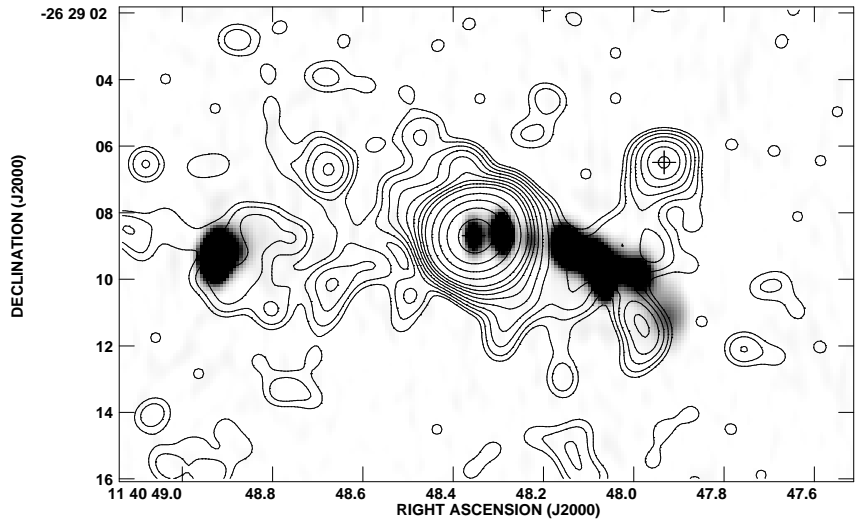}
\vskip -0.3in
\psfig{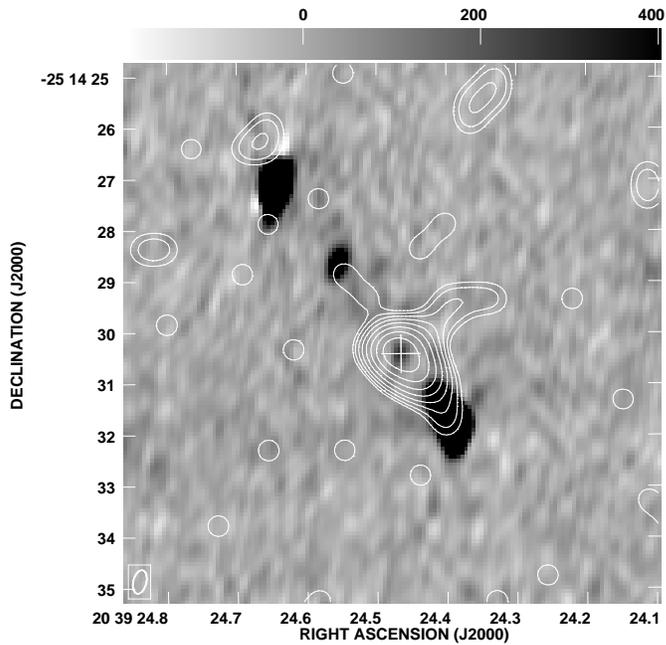}
\caption{Top: The contours show the Xray emission
from PKS 1132--262 and the greyscale shows the radio emission
at 5 GHz (Carilli et al. 2002). Bottom: The contours show
the Xray emission from 2036--254  and the greyscale shows the 
radio emission at 8 GHz.}
\end{figure}

A second mechanism is non-thermal radiation, either synchrotron or
IC. For 2036--254 (Figure 3) the Xray emission coincides with the 
inner part of the radio lobe situated closest to the nucleus
(presumably the receding radio lobe). The Xray and radio emission
show opposite gradients, with the Xrays decreasing with increasing
distance from the nucleus, while the radio increases. This morphology
is consistent with the Brunetti mechanism of IC
scattering of radiation from the AGN by the relativistic electrons in
the radio source (Brunetti 2000). 

For 2036--254 it can be shown that the energy
density in the radiation field is dominated by photons from the active
nucleus within about 10 kpc of the AGN, assuming an AGN luminosity of
10$^{46}$ erg s$^{-1}$ (as opposed to the CMB and the radio synchrotron
photons).  Moreover, Brunetti (2000) predicts that the receding
lobe will be brighter in Xrays than the approaching lobe both because
it is closer to the nucleus (due to the time delay) and because
back-scattering is more efficient than forward-scattering. 
Comparing the IC Xrays and the synchrotron radio emission implies a
magnetic field of 25 $\mu$G is the radio source, which is a factor three
below minimum energy.

The most spectacular example of the radio-Xray alignment effect is
PKS 1138--262.  This source has been considered in detail by Carilli
et al. (2002). They find that multiple mechanisms may be at work in
1138--262. Emission from the radio galaxy AGN, and a second AGN about
6$''$ to the northwest is detected.  The Xray emission from the inner
radio jet is most likely due to the Brunetti mechanism described above,
with implied magnetic fields of about 30$\mu$G.

For the outer radio jet the most plausible mechanism is thermal
emission from gas shock heated by the expanding radio source. The
implied parameters for the thermal gas are: $\rm n_e = 0.05$
cm$^{-3}$, $\rm M_{gas} = 2.5\times10^{12}$, and $\rm P_{gas} =
8\times10^{-10}$ dynes cm$^{-2}$. The pressure in the Xray emitting
gas is comparable to that in the (low filling factor) optical line
emitting gas, and to minimum pressures in the ratio source.  Chambers
et al. (1990) first pointed out the need for a high filling factor,
hot gas to confine the optical line emitting gas and the radio source
in high redshift radio galaxies.

Rees (1989) suggested that the ambient medium into which a radio jet
propagates may be very different for high redshift sources relative to
low redshift sources. For the low redshift sources the medium is a
relatively smooth, virialized (10$^8$ K) cluster atmosphere. For the
high redshift sources such a smooth medium has not had time to
develop. The ambient medium is likely to be multiphase, with
structures ranging from low filling factor, high density ($>$ few
cm$^{-3}$), cold (10$^4$ K) clouds, to high filling factor, lower
density ($< 0.01$ cm$^{-3}$) regions at the virial temperature of
10$^6$ to 10$^7$ K. The passage of a radio jet through such a
multiphase medium will lead to a number of interesting phenomena
(Kaiser \& Alexander 1999). The high density clouds are induced to
form stars due to the higher pressure environment. The intermediate
density regions are shocked, but cool on timescales comparable to the
radio source lifetime ($10^7$ years), and emit Ly $\alpha$
radiation. The lower density, higher filling factor gas is shock
heated to Xray emitting temperatures.  The Chandra observations of
1138--262 may have revealed, for the first time, this hot
and pervasive medium, as required to confine the radio
source and the optical line clouds. 

Optical observations of radio galaxies and their environments have
revealed cluster-like overdensities of galaxies (Venemans these
proceedings). However, the velocity dispersions are smaller than
expected for a virialized cluster. Likewise, the Xray observations of
high redshift radio galaxies have shown that they are not at the
center of massive, virialized cluster atmospheres, and suggest the
existence of a much less mature ambient medium. Overall, these
observations are consistent with the idea that high redshift radio
galaxies are in proto-cluster environments, ie.  regions that will
evolve into dense clusters, but have not yet had time to fully
separate from the Hubble flow and virialize.

This paper summarizes work done with many collaborators, including
H. Rottgering, D. Harris, L. Pentericci, G. Miley, R. Overzier,
J. Kurk.  The National Radio Astronomy Observatory (NRAO) is operated
by Associated Universities, Inc. under a cooperative agreement with
the National Science Foundation. This research was supported by a
grant from the Chandra observatory, and made use of the Chandra
archive.

\end{document}